\documentclass[12pt]{article}

\usepackage{amsfonts, amsmath}

\newcommand{\be}{\begin{equation}}
\newcommand{\ee}{\end{equation}}
\newcommand{\lbl}[1]{\label{#1}}

\newcommand{\vect}[1]{\ensuremath{\boldsymbol{#1}}}

\def\diag{\ensuremath{\mathop{\rm{diag}}}}
\def\ndd{\noindent}

\begin{document}

\title{On non-Adiabatic Holonomoic Quantum Computer}
\author{A.E.Margolin \hspace{1.5mm}$^1$\thanks
{Phone (+375) 172 283438; e-mail alexm@hep.by},
V.I.Strazhev$^1$
\hspace{1mm} and \hspace{1mm} A.Ya.Tregubovich$^2$
\thanks{Phone (+375) 172 283438; e-mail a.tregub@open.by}
}
\date{}
\maketitle
 \vspace{-25pt}
{\footnotesize\noindent $^1$National Centre of High Energy Physics,Bogdanovich
 Str.153, Minsk 220040, Belarus\\
$^2$Institute of Physics National Academy of Sciences
                          Skoryna av.68, Minsk 220072, Belarus}\\

{\bf\small\noindent Abstract}\\
{\footnotesize
Non-adiabatic non-Abelian geometric phase of spin-$3/2$ system in the rotating magnetic
field is considered. Explicit expression for the corresponding effective non-Abelian
gauge potential is obtained. This formula can be used for construction of quantum gates
in quantum computations.
}
\vspace{0.5cm}
{\ttfamily{\footnotesize  geometric phase}}

\rm\normalsize
\vspace{0.5cm}

A great deal of investigations in quantum computers (see \cite{cabello} for a bibliographic
review on this subject) refreshed some interest on Berry phase effect \cite{berry}
in quantum mechanics. The idea of using unitary evolution operators produced by
the non-Abelian Berry phase \cite{wilczek} as quantum calculations is proposed in
\cite{zanardi} and realized in \cite{pachos} in a concrete optical model of holonomic
quantum computer, in \cite{choi} in the solid state model, in the model of network
of Josefson junctions \cite{faoro}, in \cite{pachos2} for the ion-trap model
 and in \cite{recati} for that of QED cavity.
Calculation aspects of this model are considered in
\cite{margolin}. For other references where also Abelian Berry phase is considered
in the context of quantum computer see e.g. \cite{ekert} - \cite{averin}.
  On the other hand non-adiabatical Berry phase can exist and be
measured  if transitions in a given statistical ensemble do not lead to loose of
coherency \cite{appelt}. Therefore it is also possible to use the corresponding
unitary operators to perform quantum calculations. This fact has been noticed
in \cite{keiji}, \cite{blais} for the NMR model of QC and in \cite{li} for the
 ion-trap model. In this paper we show a realization of quantum gates for a
concrete 4-level quantum system driven by external magnetic field.
Let us consider a spin-$3/2$ system with quadrupole interaction.
Physically it can be thought of as a single nucleus with the spin above. A
coherent ensemble of such nuclei manifest geometric phase when placed in
rotating magnetic field. This phase is non-Abelian due to degenerate energy
levels with respect to the sign of the spin projection. Depending on the
experiment setup the phase can be both adiabatic as in Rb experiment by Tycko
\cite{tycko} and non-adiabatic as in Xe experiment by Appelt et al
\cite{appelt}. This non- Abelian phase  results in mixing of $\pm 1/2$ states
in one subspace and $\pm 3/2$ in another one and thus can be regarded as a
2-qubit gate. The gate is generated by a non-Abelian effective gauge potential
${\bf A}$  that is the subject of computation in this paper.\\
\ndd
Let us consider a parametric quantum system described by a Hamiltonian
$H({\vect{R}})$, where $\vect{R}(t)$ is a set of evolving parameters. We suppose
that evolution of the Hamiltonian is determined by a unitary rotation
\begin{equation}\label{unirot}
  H(t) = U^{\dagger}(t)\, H_0(t)\,U(t) ,
\end{equation}
where $H_0(t)$ is diagonal in some basis in the Hilbert space. Looking
for solutions of the Schr{\"o}dinger equation
\begin{equation}\label{schr1}
  i{\partial \psi \over\partial t} = H(t)\, \psi (t),
\end{equation}
we take rotating frame by assigning $\tilde{\psi} (t) = U(t)\,\psi(t)$ and get
in such a way
\begin{equation}\label{schr2}
  i{\partial \tilde{\psi} \over\partial t} = (H_0(t) - i\,U^{\dagger}(t)\dot{U}(t) )
  \, \tilde{\psi}(t).
\end{equation}
Of course, this transformation generally does not help to solve equation
(\ref{schr1}) due to the fact that the algebraic structure of the coupling term
$- i\,U^{\dagger}\dot{U}$ can appear to be rather complicated and the last
generally does not commute with $H_0$. However if a receipt is known how to
evaluate the last term in (\ref{schr2}), further solution of this equation is
straightforward:
\begin{equation}\label{gensol}
   \psi (t) = e^{-i\,\phi_n(t)}\;\mathcal{T}\!\exp\left(-i\,\int_0^t\,
    U^{\dagger}(\tau)\dot{U}(\tau)\, d\tau \right)\, \psi (0) ,
\end{equation}
where $\phi_n(t) = \int_0^t\,E_n(\tau)\, d\tau$ is so called dynamical phase,
$\mathcal{T}$ denotes time-ordering and $E_n$ are elements of $H_0$ that is by
definition diagonal. Of course if there is no way to find
$U^{\dagger}(\tau)\dot{U}(\tau)$, expression (\ref{gensol}) is useless, but
there is at least one sufficiently general case in which this term can be
independently computed. It is the case of a parametric quantum system with
adiabatically evolving parameters. The corresponding class of solutions was discovered by
Berry \cite{berry} for non-degenerate $E_n$ and was generalized by Wilczek and
Zee \cite{wilczek} for the degenerate case. In this case
\begin{equation}\label{adiab}
  (U^{\dagger}(t)\dot{U}(t))_{ab} =  (\vect{A}_n)_{ab}(\vect{R}) =
  -i\,<\varphi^{(n)}_b|\nabla_{\vect{R}}\,\varphi^{(n)}_a> ,
\end{equation}
where $\vect{R} = (R_1,...,R_N)$ is a set of externally driven parameters and
$\varphi^{(n)}_a\quad a=1,...,\kappa_n$ ($\kappa_n$ is the degeneration rate
for the energy level $E_n$) are solutions of the eigenvalue problem
$$
     \hat{H}_0(\vect{r}, \vect{R}(t))\,\varphi^{(n)}_a =
     E_n(\vect{R}(t))\,  \varphi^{(n)}_a(\vect{r}, \vect{R}(t)).
$$
Here $\vect{r}$ denotes the set of dynamic variables affected by $\hat{H}_0$.
To our knowledge other classes of quantum systems that admit such a transparent
evaluation of the term $U^{\dagger}\dot{U}$. In this paper we consider a
particular case of 4-level system investigated experimentally in \cite{appelt}.
For this system a constructive derivation of the above term is possible.
Evolution of the magnetic field governing the the system can be generally
non-adiabatic so the solution is of more general type (\ref{gensol}) but not of
(\ref{adiab}). It is important to realize that although the solution part corresponding
to the second factor in the right hand side of formula (\ref{gensol}) can also
 be called "geometric phase", it
can not be computed by means of formula (\ref{adiab}) because of the system's
non-adiabaticity.

 We assume
the condition of the $^{131}Xe$ NMR experiment to be held so one does not need
 to trouble about the coherency in the system.
The last is described by
the following Hamiltonian in the frame where the magnetic field is parallel
to the z-axis ($\hbar = 1$)
\be\lbl{pokham}
H_0 = \omega_0 (J_3^2 - 1/3 j(j+1))
\end{equation}
 For a spin-$3/2$ system we choose the third projection of the angular momentum in the
form
\be
J_3 = \begin{pmatrix}
                        3/2  &  0  &  0  &  0  \\
                         0   &-3/2 &  0  &  0  \\
                         0   &  0  & 1/2 &  0  \\
                         0   &  0  &  0  &-1/2  \\  \end{pmatrix} =
  \begin{pmatrix}
   3/2 \sigma_3 & 0 \\
    0           & 1/2 \sigma_3 \\
  \end{pmatrix},
\end{equation}
Then two other projection operators are \be \hspace*{-24pt} J_1 =
\begin{pmatrix}
    0             &  0        &  \sqrt{3}/2  &  0           \\
    0             &  0        &       0      & \sqrt{3}/2 \\
\sqrt{3}/2        &  0        &       0      &  1           \\
     0            & \sqrt{3}/2&       1      &  0          \\
\end{pmatrix} =
  \begin{pmatrix}
     0     & \sqrt{3}\over 2 \\
\sqrt{3}\over 2 & \sigma_1 \\
  \end{pmatrix},
\end{equation}
\be
\hspace*{12pt}
    J_2 = \begin{pmatrix}
        0             &  0        &       0      &  -\sqrt{3}/2  \\
        0             &  0        & \sqrt{3}/2   &       0       \\
        0             &\sqrt{3}/2 &      0       &      -i        \\
      -\sqrt{3}/2     &    0      &      i       &       0        \\
    \end{pmatrix} =
   \begin{pmatrix}
      0               & -i{\sqrt{3}\over 2}\, \sigma_3  \\
i{\sqrt{3}\over 2}\, \sigma_3  & \sigma_2               \\
   \end{pmatrix}.
\end{equation}
In the laboratory frame the Hamiltonian takes the form
\begin{equation}\label{movHam}
  H = \omega_0 ((\vect{J}\vect{n})^2 - 1/3 j(j+1)) =
       e^{-i\varphi J_3}  e^{-i\theta J_2}\, H_0 \,
       e^{i\varphi J_2}  e^{i\theta J_3}.
\end{equation}
 Rotation around the z-axis means that $\varphi =\omega_1 t$ and one should
 perform the unitary transformation

\begin{equation}\label{u1}
  |\psi > = U_1\, |\tilde\psi > ,\qquad U_1 = e^{-i\omega_1 t J_3}.
\end{equation} In the rotating frame we get
\begin{equation}\label{h1}
  H_1 = e^{-i\theta J_2}(\omega_0 J_3^2 - \omega_1\tilde J_3)e^{i\theta J_2} -
        {5\omega_0\over 4},
\end{equation}
where
$$
      \tilde J_3 = e^{i\theta J_2}\, J_3 \, e^{-i\theta J_2}
$$ Expression (\ref{h1}) is equivalent to

\begin{equation}\label{h1matr}
  H_1 = \begin{pmatrix}
         \omega_0 - {3\over 2}\omega_1 \cos\theta\,\sigma_3 &
                  \omega_1{\sqrt{3}\over 2} \\
         \omega_1{\sqrt{3}\over 2}            &  -\omega_0 - {1\over 2}\omega_1
                                \cos\theta\,\sigma_3  +
                                \omega_1\sin\theta\,\sigma_1\\
        \end{pmatrix}
\end{equation} It is convenient to diagonalize this matrix in two steps. First
we get rid of $\sigma_1$ in the last matrix element by performing of the
block-diagonal transformation
\begin{equation}\label{u2}
  U_2 = \diag (1, e^{-i\alpha\,\sigma_3}),
\end{equation}
where $\tan\alpha = 2\tan\theta$. Thereafter the Hamiltonian $H_1$ reads
\begin{equation}\label{h2}
 H_1 = \begin{pmatrix}
         \omega_0 - {3\over 2}\omega_1 \cos\theta\,\sigma_3 &
                  \omega_1{\sqrt{3}\over 2} \\
         \omega_1{\sqrt{3}\over 2}            &  -\omega_0 - {1\over 2}\omega_1
                                {\cos\theta\over  \cos\alpha}\,\sigma_3  \\
        \end{pmatrix}.
\end{equation}
At the second step we apply the transformation
\begin{equation}\label{u3}
 \begin{pmatrix}  \beta_1  & \beta_2\\
                -\beta_2^* & \beta_1^*\\
    \end{pmatrix},
\end{equation}
where $\beta_1$, $\beta_2$ are diagonal $2\times 2$ matrices that must obey the
unitarity condition
\begin{equation}\label{cnd1}
  |\beta_1|^2 + |\beta_2|^2 = 1.
\end{equation}
Supposing $\beta_{1,2}$ to be real and performing transformation (\ref{u3}) we
come to the diagonalization condition in the form
\begin{equation}\label{cnd2}
  \xi (\beta_1^2 - \beta_2^2) + (\lambda_1 - \lambda_2)\beta_1\beta_2 = 0,
\end{equation}
where $\lambda_1, \lambda_2$  are $2\times 2$ diagonal matrices and $\xi$ is
a parameter
\begin{eqnarray}\label{lmd12}
\lambda_1 &=& \omega_0 - 3/2\,\omega_1 \cos\theta\,\sigma_3, \\
\lambda_2 &=& -\omega_0 - 1/2\,\omega_1
                               {\cos\theta\over  \cos\alpha}\,\sigma_3\\
\xi &=& \omega_1\,\sqrt{3}/2\,\sin\theta
\end{eqnarray}
Assuming $\beta_2 = \mu\,\beta_1$ where $\mu$ is a diagonal $2\times 2$ matrix
as well we come to the following expressions for the matrix elements of $\mu$
\begin{equation}\label{mui}
  \mu_i = k_i + \sqrt{1 + k_i^2},
\end{equation}
where
$$
k_i = {\Delta\lambda_i\over 2\xi}, \quad
\Delta\lambda_i = \lambda_{1i} - \lambda_{2i}.
$$
Finally we get for the matrix elements of $\beta_{1,2}$
\begin{eqnarray}\label{beta12}
  \beta_{1i}^2 &=& 1/2(1 + k_i^2)^{-1/2}\left(k_i + \sqrt{1 + k_i^2}\right)^{-1}\\
  \beta_{2i}^2 &=& 1/2\left(1 + {k_i\over\sqrt{1 + k_i^2}}\right)
\end{eqnarray}
Now one can evaluate the connection 1-form. It is convenient to represent it
as follows:
\begin{equation}\label{A}
 \vect{A} = i\, U^+dU = A\, d\phi = \begin{pmatrix}
            A_{3/2}          & A^{tr}\\
            {\tilde A^{tr}}  & A_{1/2}  \end{pmatrix}\, d\phi,
\end{equation}
where all matrix elements of $A$ denote $2\times 2$ matrix-valued blocks,
$U = U_1U_2U_3$ and $U_i$ are determined by (\ref{u1}), (\ref{u2}),
(\ref{u3}) correspondingly. Here tilde denotes a transposed matrix.
After some algebra we get for the matrix elements of (\ref{A})
\begin{eqnarray}\label{Aij}
  A^{tr}  &=& {1\over 2} \beta_1\beta_2(3 - \cos\alpha)\sigma_3
   + {1\over 2}\sin\alpha\,\beta_2\sigma_1\beta_1 , \\
  A_{3/2} &=& (a_{3/2} + b_{3/2}\,\sigma_3 + c_{3/2}\,\sigma_1)\, d\phi, \\
  a_{3/2} &=& {1\over 4} \left(3\beta_{11}^2 - 3\beta_{12}^2 +
             \beta_{21}^2\cos\alpha - \beta_{22}^2\cos\alpha\right), \\
  b_{3/2} &=& {1\over 4} \left(3\beta_{11}^2 + 3\beta_{12}^2 +
             \beta_{21}^2\cos\alpha + \beta_{22}^2\cos\alpha\right), \\
  c_{3/2} &=& - {1\over 2}\,\sin\alpha\,\beta_{21}\beta_{22},\\
  A_{1/2} &=& (a_{1/2} + b_{1/2}\,\sigma_3 + c_{1/2}\,\sigma_1)\, d\phi, \\
  a_{1/2} &=& {1\over 4} \left(3\beta_{21}^2 - 3\beta_{22}^2 +
               \beta_{11}^2\cos\alpha - \beta_{12}^2\cos\alpha\right), \\
  b_{1/2} &=& {1\over 4} \left(3\beta_{21}^2 + 3\beta_{22}^2 +
               \beta_{11}^2\cos\alpha + \beta_{12}^2\cos\alpha\right), \\
  c_{1/2} &=& - {1\over 2}\,\sin\alpha\,\beta_{11}\beta_{12},
\end{eqnarray}
where $d\phi = \omega_1 dt$. Note that as $A$ does not depend on time, the final
solution does not require $\mathcal{T}$-ordering. It should be also emphasized
 here that in the
non-adiabatic case we discuss the term $A_{3/2}$ contains non-diagonal terms
that is not the case when the adiabaticity condition is held \cite{moody}.
Now the solution of the problem takes a particular form of (\ref{gensol}):
\begin{equation}\label{partsol}
   \psi (t) = e^{-i\,\phi_n(t)}\; e^{-i\omega_1 t\, A}\; \psi (0) ,
\end{equation}

Formula (\ref{partsol}) solves the problem of the evolution control for the system
under consideration. The resulting quantum gate is entirely determined by
$ A$ and the evolution law of the magnetic field, i.e. by a contour in the
parameter space. Of course it is always possible to choose the parameters so that
$A$ turns out to generate a 2-qubit transformation that produces a superposition
of basis states. For this reason
the gate can be thought of as a universal one \cite{deutsch}.
 Of course, a suitable speed of the parameters evolution
can not be reached by rotation of the sample as it took place in the experiment
by authors of \cite{appelt}. Nevertheless it is clear that this manner of control
is not principle and one could imagine a situation where the parameters evolution
is provided by the controlling magnetic field by adding a non-stationary transverse
component.

\end{document}